\UseRawInputEncoding
\documentclass[conference, 10pt]{IEEEtran}
\usepackage{amssymb,amsmath}
\usepackage{amssymb,amsmath}
\usepackage{setspace}
\usepackage{graphicx}
\usepackage{epsfig,cite,latexsym,subfigure}
\usepackage{theorem}
\usepackage{times}
\usepackage{epsf,cite,subfigure}
\usepackage{amsmath,amssymb}

\usepackage{floatflt}
\usepackage{url}
\usepackage{times}
\usepackage[usenames]{color}
\usepackage{dsfont}
\usepackage{mathptmx}
\usepackage[final]{pdfpages}
\newcommand{\set}[1]{\left[#1\right]}
\newcommand     {\curlb}[1]{\left\{ #1\right\} }
\newcommand     {\paren}[1]{\left(#1\right)}
\newcommand{\eqnlabel}[1]{\label{eqn:#1}}
\newcommand{\eqnref}[1]{(\ref{eqn:#1})}
\DeclareMathOperator*{\argmax}{arg\,max}

\title{{\Large Practical Fingerprinting of RF Devices in the Wild}}\vspace{-2mm}
\author{
\IEEEauthorblockN{\normalsize Silvija Kokalj-Filipovic, Luke Boegner, Robert D. Miller} \\ \vspace{-3mm}\\
\IEEEauthorblockA{\small Perspecta Labs Inc. \\
\small\em \{skfilipovic, luke.boegner, rmiller\}@perspectalabs.com}}%, 
\begin{document}
\maketitle
\begin{abstract}
We present a new RF fingerprinting technique for wireless emitters that is based on a simple, easily and efficiently retrainable Ridge Regression (RR) classifier.  The RR learns to identify devices using bursts of waveform samples, conveniently transformed and preprocessed by delay-loop reservoirs. Deep delay Loop Reservoir Computing (DLR) is our processing architecture that supports general machine learning algorithms on resource-constrained devices by leveraging delay-loop reservoir computing (RC) and innovative architectures of loop trees.  In prior work, we trained and evaluated DLR using high SNR device emissions in clean channels.  We here demonstrate how to use DLR for IoT authentication by performing RF-based Specific Emitter Identification (SEI), even in the presence of fading channels and heavy in-band jamming by leveraging a matched filter (MF) extension, dubbed MF-DLR. We show that the MF processing improves the SEI performance of RR without the RC transformation (MF-RR), but the MF-DLR is more robust and applicable for addressing signatures beyond waveform transients (e.g. turn-on).
\end{abstract}
%\vspace{-2mm}
%\section*{}
\section{Introduction}%\vspace{-1mm}
The Internet of Things (IoT) is becoming a pervasive reality.  Billions of IoT devices will be deployed with vastly different computational capabilities, however they must all be protected.  Conventional authentication mechanisms to verify device identities are based on upper layer protocols that may be too complex and resource-restrictive for use on many IoT devices.  Such techniques tend to be cryptographically-based and require active collaboration, hence involving computational resources, power drain, and may still not be able to meet latency requirements. Physical layer authentication (PLA) methods are relatively new \cite{WiPLI}, and the research \cite{PLASurvey} mostly addressed 2 types of  PLA approaches: 1) using Channel State Information (CSI) 2) exploiting the HW signature of the transmitting device in the received signal. Both methods eliminate the need for cryptography and the associated overhead,  with the 2nd approach generally more reliable. As all electronic devices have fingerprints due to manufacturing variability, so do radio frequency emitters. The composition of these RF fingerprints is described in detail in \cite{WiPLI}: it includes, among others things, clock jitter, DAC sampling inaccuracy, mixer or local frequency synthesizer imperfections, power amplifier non-linearity, device antenna and  modulator sub-circuit (if the analog modulation is used). A convenient name  for such fingerprints used in emitter identification was coined in \cite{PUF}: Physical unclonable functions (PUFs). PUF-based systems for unique identification of emitters are extremely robust and secure at low cost, as it is practically impossible to replicate the same silicon characteristics across dies. Specific Emitter Identification (SEI) based on RF fingerprinting can individually identify wireless devices, including previously unseen \cite{CabricSEIAuthor}.  Even though  RF fingerprints have been analyzed and modeled earlier \cite{WiPLI}, their full potential for SEI became clear with deep learning.  The most recent work is on the SEI designs based on supervised deep learning \cite{Merchant}. This approach trains neural network models to verify the identity of wireless transmitters using labeled datapoints derived from the captured emissions of those transmitters without any feature engineering. 
%%%%SKF - TODO %%%

Our approach to SEI for IoT authentication is based on leveraging signatures present in the turn-on transients of IoT emissions.  Contrary to what is stated in \cite{PUF}, in combination with our resource-constrained but extremely efficient classifier, these signatures prove to be accurate and robust. Our SEI design based on delay-loop reservoirs  
used a sampling rate of 100Msps as opposed  to the examples featured in  \cite{PUF} (500 Msps in \cite{BTRFfing} and 50
GS/s in \cite{transRFfing}. Hence, the expensive receiver architectures are no longer required.
 
\begin{figure}[h]
\vspace{-5mm}
\centering
\hspace{-1mm}\includegraphics[width=0.5\textwidth]{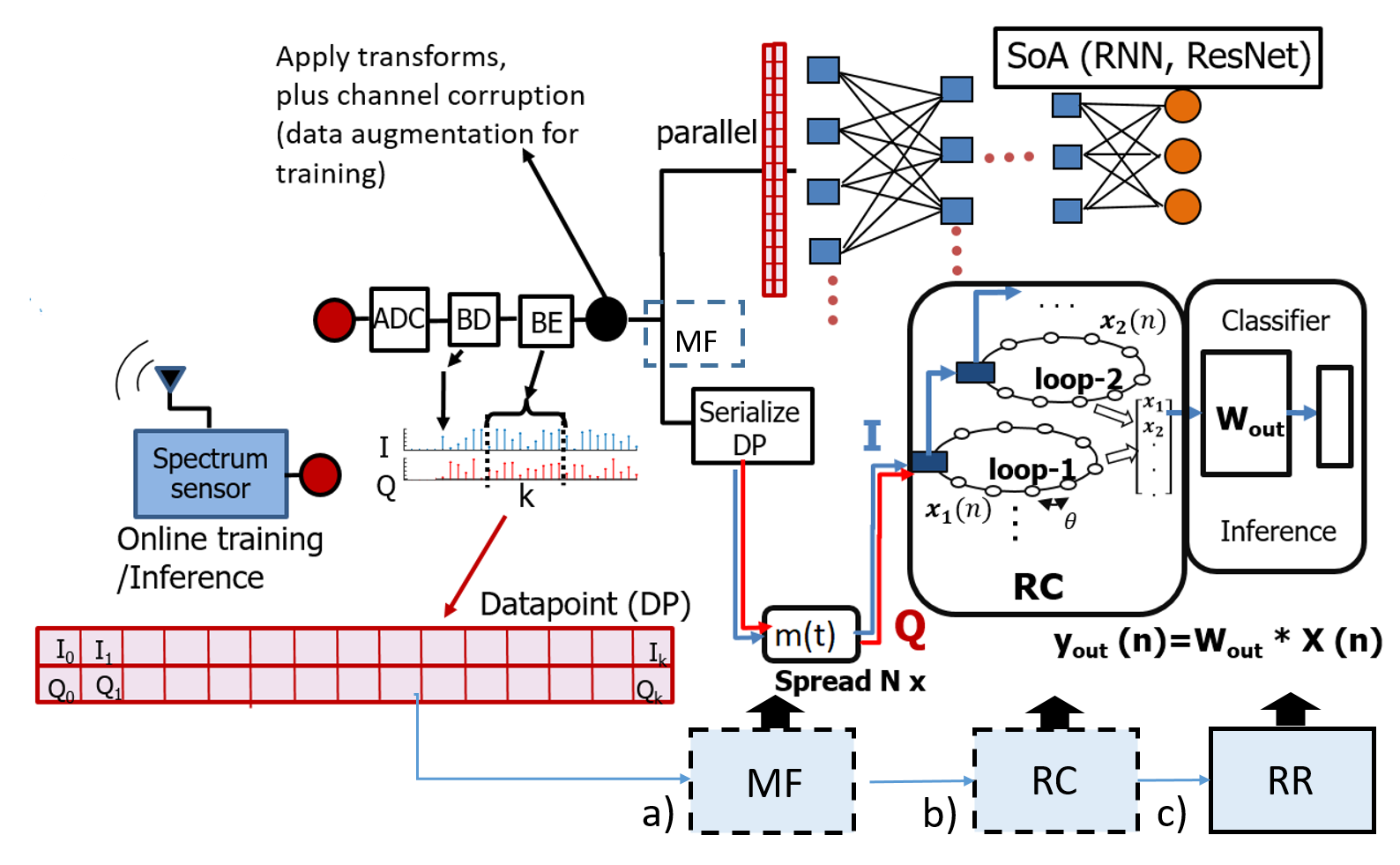}
\vspace{-4mm}
\caption{DLR system for SEI/WiPRec: BD is burst detection, while BE stands for burst extraction. {\bf DLR} includes blocks $b + c,$ {\bf MF-DLR} includes $a + b + c$, where $c$ is the simple {\bf RR} classifier.
}\vspace{-3mm}
\label{fig:DLRsys}
\end{figure}

%%[28] I.O. Kennedy, P. Scanlon and M.M. Buddhikot, “Passive Steady State RF
%%Fingerprinting: A Cognitive Technique for Scalable Deployment of CoChannel Femto Cell Underlays,” IEEE Symposium on New Frontiers in
%%Dynamic Spectrum Access Networks, 2008.
%%[29] B. Kroon, S. Bergin, I.O. Kennedy and G. Zamora, “Steady State RF
%%Fingerprinting for Identity Verification: One Class Classifier versus
%%Customized Ensemble,” Irish conference on Artificial Intelligence and
%%Cognitive Science (AICS), 2009.
%%[30] K. Merchant, S. Revay, G. Stantchev and B. Nousain, “Deep Learning for
%%RF Device Fingerprinting in Cognitive Communication Networks,” IEEE
%%Journal of Selected Topics in Signal Processing, 2018.
%%[31] T.J. Bihl, K.W. Bauer and M. A. Temple, “Feature Selection for RF
%%Fingerprinting With Multiple Discriminant Analysis and Using ZigBee
%%Device Emissions,” IEEE Transactions on Information Forensics and
%%Security, 2016. 	
Further, our solution to efficient RF fingerprinting of wireless emitters is based on a simple, easily retrainable classifier that learns to identify emitting devices based on bursts of waveform samples, conveniently transformed and preprocessed by a delay-loop reservoir.  State-of-the-Art (SoA) machine learning systems that are trained on sensor signals lack the computational resources to support in-situ training and adaptable inference for situational awareness. In our prior work \cite{GomacTech, reservoirJSAC}, we proposed a solution through Deep delay Loop Reservoir Computing (DLR), our novel AI processing architecture that supports general retrainable machine learning (ML) solutions on compact mobile devices by leveraging delay-loop reservoir computing (RC). Reservoir computing is a bio-inspired approach suited for processing time-dependent information in a computationally efficient way \cite{Lukosevicius2009ReservoirCA}. The RC in ML solutions conditions the input features towards linear separability of different classes, upon which any ML algorithm can be trained more efficiently.  Upon transforming the inputs by delay loops, we can classify fingerprints by employing a classifier based on Ridge Regression trained on delay loop outputs. DLR delivers significant reductions in form factor and power consumption for training at the Edge, providing real-time learning latency.  For a pictorial illustration of the process employing a DLR system to train and perform SEI, including RF spectrum sensing and preprocessing, please consult  Fig.~\ref{fig:DLRsys}. %, borrowed from our paper \cite{GomacTech}, which also reports intermediate results of our research. 
Our innovative architecture of split loops yielded a resource-savvy SEI solution,  which  for a clear channel delivery of emitted signals (high SNR) exceeds the SoA accuracy.   
In this paper we demonstrate how to use DLR for IoT authentication by performing RF based Specific Emitter Identification (SEI) even in the presence of fading channels, heavy in-band jamming and receiver imperfections. In order to make SEI robust to these imperfections, we propose to use a matched filter (MF) extension to DLR where the filter is designed by processing the transient turn-on part of the device emission. We refer to it as {\bf MF-DLR}. It increases the SEI resilience to in-band jamming, channel and receiver imperfections.

We also show that the addition of the MF based on turn-on signatures improves the SEI performance of the simple Ridge Regression (RR) classifier with no delay-loop processing (parts {\bf a and c} in Fig.~\ref{fig:DLRsys}). Although MF-DLR  demonstrates better robustness to signal impairments, the improved MF-RR performance is an important observation given the simplicity and pervasiveness of regularized least square methods (RR). It appears that the MF contributes to linear separability of the signature classes, while the RC increases the margins around hyperplanes.
The organization of this paper is as follows: Section II introduces the DLR for SEI, and its components. Section III first describes DLR performance on clean data, then introduces MF-DLR and MF-RR and evaluates their robustness to various channel and receiver imperfections compared to DLR.
%%%%%%%%%%%%%%%%%%%%%%%%%%%%%%%%%%%%
%%%%%%%%%%%%%%%%%%%%%%%%%%%%%%%%%%%%%%%
 
%
%\section{SEI Problem Definition}
%%%%%%%%%%%%%%%%%%%%%%%%%%%%%%%%%%%%%%%%%%%%%%%%%
\section{DLR for SEI: System Description} 
\subsection{Basic DLR}
\begin{figure}[h]
\vspace{-1mm}
\centering
\includegraphics[width=0.46\textwidth]{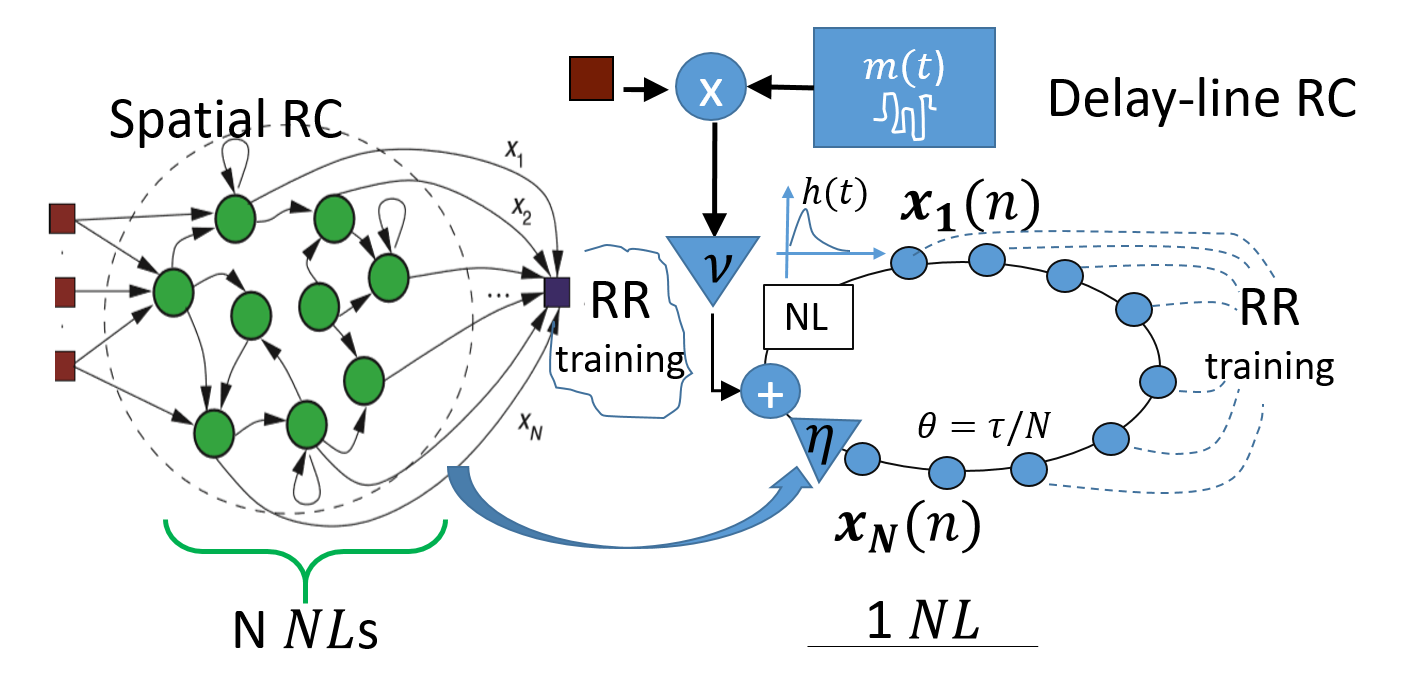}
\vspace{-5mm}
\caption{Delay-loop reservoir replaces the N neurons (NLs) on the left with a single one by using upsampling ($\theta$ depends on the loop bandwidth) and a simple design shown on the right}\vspace{-5mm}
\label{fig:basicdl}
\end{figure}

DLR uses reservoir computing in the form of a delay loop reservoir that replaces the N neurons in the traditional spatial implementation of the reservoir with N passes through a single neuron. The N-fold increase in delay by the sequential passing of the data through a single neuron is canceled out by an N-fold upsampling (sample and hold) done by a random spreading sequence $m(t)$ (see Fig.~\ref{fig:basicdl}). 
%%%%%%%%%%%%%%%%%%%%%%%%%%%%%%%%%%%%
With reservoirs, the N must be large and such high rate upsampling is amenable to a photonic loop implementation. The wide bandwidth and wavelength diversity of photonic loops is the reason why DLR supports photonic DLs as a vehicle to future scalability. We here limit the scope to the FPGA and S/W (digital) implementations of the loop.%, as the emphasis is on algorithmic and architectural solutions. 

The N which allows the reservoir to linearly separate input classes for easier learning may be very large for certain applications. However, in our prior work \cite{reservoirJSAC} we introduced algorithms for reducing the required reservoir size by means of split loops. Once trained, the DLR maps the signal to its unique transmitter by just passing the burst of samples through the loop and multiplying it with the weight matrix. For a burst of 1 us, it happens in sub-millisecond including the extraction of the burst and its preprocessing by transforms.

{\bf Dataset and Preprocessing~}
The dataset to train the SEI detector contains 20 classes, corresponding to 20 distinct WiFi devices. We captured the emissions of commercial WiFi devices as they were sending beacons to an access point while using the same spoofed MAC address. 
We used USRP X310 with UBX RF daughterboard, with sample rate of 100 MHz centered in the middle of the 2.4 GHz ISM band. Datapoints are created by extracting bursts of 1024 complex (I/Q) samples from the captured time-series, right after the detection of the rising edge of the signal. Further processing steps are explained below. 

{\bf Delay Loop (DL)~}
%The basic algorithm for the delay-loop state is expressed by%~\eqnref{delayloop}.  
%\begin{align}
%\nonumber X_k (n)=\int_{\delta_k-\tau-\epsilon}^{\delta_k}{h_{\delta_k-\tau-\delta} f_{NL} \set{\eta X_\delta (n-1)+\nu J(n)(\delta + \tau)}d\delta.}  %\eqnlabel{delayloop}
%\end{align}
Each sample $s(n), n\in{1,\cdots,\ell}$ of the input datapoint of size $\ell$ (Fig.~\ref{fig:DLRsys}) is spread by the mask $m(t)$ and clocked into the loop as input $J(n)(t),$ chip-by-chip. The upsampled time is defined in chips $\theta$ (in Fig.~\ref{fig:basicdl}).  Here, $t \in {1,\cdots,N}$ is the chip-time index, and $k \in {1,\cdots,N}$ is the loop position index. $X_k$ is the $k^{th}$ virtual node of the state vector $X.$ Note that $N$ is the number of virtual reservoir nodes, as well as the length (in chips) of the random spreading sequence (mask) $m(t),$ as in Fig.~\ref{fig:basicdl}.  The loop output is read out after the last of the $\ell$ samples is clocked-in and put through the loop's non-linearity $f_{NL}$ $N$ times.  

Each chip of the spread sample $J(n)(t)$ is linearly combined with the tail of $X: X_N(n-1)$ and put through the nonlinearity (NL).  $X_N(n-1)$ was affected by the same non-linearity at time $t-\tau,$ i.e., by the previous input sample $n-1$, where $\tau = N\theta$. Summation of the spread data input and  the tail of $X$ at every $t$ is practically creating the edges of the recurrent layer from the spatial implementation of the reservoir (the left side of Fig.~\ref{fig:basicdl})  \cite{AdvancesinphotonicRC}.  The output of the $NL$ may be convolved with filter $h(t)$  to modify the adjacency graph of virtual nodes resulting in a faster-mixing reservoir. 
Omitting the effect of $h(t)$, $X_k$s are simply time-shifted, in chip time $\theta,$ which also roughly matches the propagation time in photonics. \eqnref{delayloopD} models the shifting of the $f_{NL}$ output in digital implementation, where $X_k$ at chip time $t$ is given by $X_k (t)=$
\begin{align}
%\nonumber &X_k (t)=\\
\vspace{-1mm}
 &\sum_{u=0}^{1}{h(u) f_{NL} \set{\eta X_k(t-N +u)+\nu J(t-k-u)}}  + \sigma, 
\eqnlabel{delayloopD}
\vspace{-1mm}
\end{align}
$\sigma \longrightarrow 0.$ The loop gain $\eta$  and input gain $\nu,$ as well as the taps of $h(t),$ must be calibrated to provide a proper dynamic state of the reservoir. 
The results are based on $f_{NL}=sin\paren{\cdot}.$  
%%%%%%%%%%%%%%%%%%%%%%%%%%%%%%%%%%%%
%\begin{figure}[h]
%\vspace{-1mm}
%\centering
%\includegraphics[width=0.39\textwidth]{sequentLoop1.png}
%\vspace{-4mm}
%\caption{Pseudo code for the delay loop}\vspace{-3mm}
%\label{fig:pseudo}
%\end{figure}

Splitting the input into $k$ parallel loops and combining their outputs reduces the complexity of the classifier by $k^2$ (please see the section on {\bf split loops} below). We use transforms to adapt the salient signal information to the split loop input. We develop multiple ways to merge the outputs of the constituent DLR loops into a single output to be used to train a simple RR algorithm \cite{RR}. 
We should emphasize that the loops are never trained, only the RR algorithm, which uses their outputs.
Note that the computational complexity of training the DLR (Fig.~\ref{fig:fom}) is the complexity of the last stage (RR), given the simple implementation of the loop.

{\bf Ridge Regression~}
The Ridge Regression algorithm for the estimation of the weight coefficient matrix $W$ can be presented in closed form as
$$W = (X^T X +\lambda I_N )^{-1}  \ (X^T Y_{out} )$$
where $X$ is the matrix of the $B$ state vectors used for its training,  $Y_{out}$ is the corresponding matrix of device-labels in onehot representation, and $\lambda$ is the regularization factor. Once the coefficient are trained on $X,$ the identity $D_T$ of a burst of samples $s_T(n), 0 \leq n\leq \ell$ can be determined by passing it through the reservoir $X_T = RC(s_T),$ and multiplying it with $W:$ $D_T = \argmax(X_TW).$ 

%%%%%%%%%%%%%%%%%%%%%%%%%%%%%%%%%%%%
\begin{figure}[h]
\vspace{-2mm}
\centering
\includegraphics[width=0.33\textwidth]{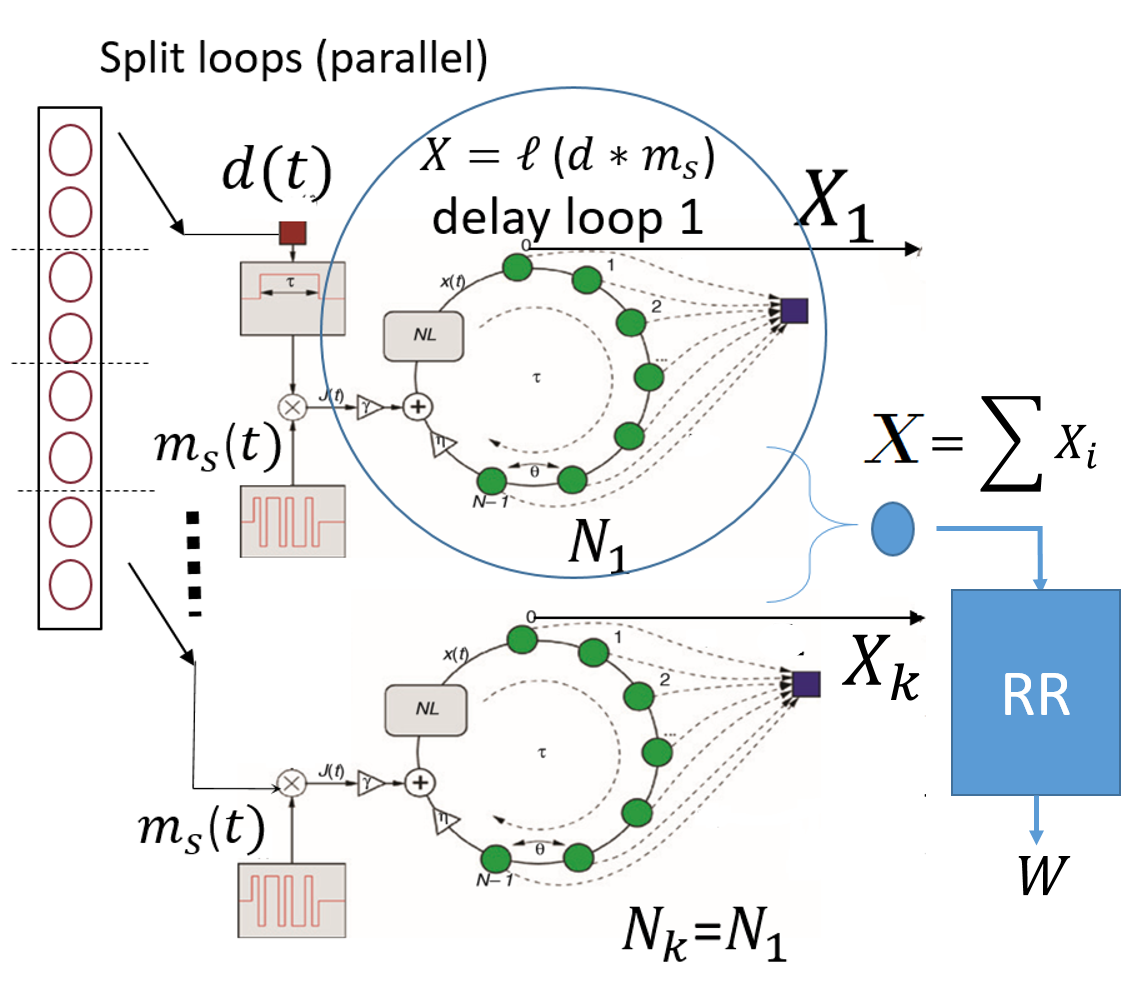}
\vspace{-6mm}
\caption{Using k split (parallel loops) reduces the size of the spreading mask and, hence, the size of each split loop while achieving the required projection into higher dimensional space. The joint state vector $X$ (marked by the blue circle) can be obtained as $\sum_{j=1}^{k}{X_j}$, or as a normalized scalar product of $X_j$s.}\vspace{-3mm}
\label{fig:multiple}
\end{figure}
%%%%%%%%%%%%%%%%%%%%
{\bf Split Loops} 
As Fig.~\ref{fig:multiple} shows, the split loops process the $k$ disjoint pieces of the split datapoint in parallel and result in the k state vectors $\curlb{X_1,\cdots, X_k}$. The joint state vector $X$ (marked by the blue circle) can be obtained as $\sum_{j=1}^{k}{X_j}$, or as a normalized scalar product of $X_j$s. Both gave similar results in terms of the improved accuracy over a single loop. Let us emphasize here that DL is a dynamic system, and the reservoir size $N$ for a datapoint of size $\ell$ must satisfy $N>>\ell$ to bring the DL into a dynamic state where the class separation happens. For spatial reservoirs, this is achieved by optimizing spectral properties of the adjacency matrix of size $N.$ With delay loops, we instead optimize by modifying $N, \eta, \nu, \lambda$ and $h(t).$
Each of the split-datapoints is of size $\ell/k$ and its $N_j<N, j=1,\cdots,k$ now may be $k$ times smaller than the $N$ without splitting. Now, as the total $X$ is a linear combination and not concatenation of the split loops' $X_j$s, the entry to the RR is still of size $N_j$, which reduces the RR complexity from $BN^2$ to  $B(N/k)^2$. Beyond a certain value of $k,$ the accuracy starts to drop as splitting the datapoint affects the samples that are no longer independent. The exact value of the threshold $k$ depends on the input transform applied to the burst of $I/Q$ samples composing the dataset. As an example, this threshold for the FFT is 10.   

{\bf Input Transforms~}
It is known that certain transforms losslessly compress the data if the information content is sparse, and some are more robust to noise, including the noise in the photonic loop. Motivated by this, we experimented with multiple input transforms in order to best mach the input to the architecture of DLR, given constraints in the complexity (i.e., largest reservoir output we are willing to train). 
%Since the RF waveforms are characterized by amplitude and phase we have 2 dimensions per sample in each 1024-long datapoint. Note that DLR loops cannot process complex-valued datapoints, and therefore all the transforms that we apply lose the phase information. 

We consider the transform from complex-valued samples to their amplitudes as the baseline. The accuracy of DLR based on the amplitudes ranks the worst among the tested transforms. We also evaluated 1) reservoir inputs made of frequency estimates \cite{SKay} of the complex valued input bursts, 2) amplitudes of the FFTs of input bursts 3) inputs made of the so called differential FFT and 3) inputs based on decimated DFT transforms. In what follows, we will use the FFT amplitudes, based on the detailed discussion in \cite{ICCpaper, reservoirJSAC}. We will use 10 splits of the FFT transform, as the FFT showed increased accuracy with number of splits $k$ up to $k=10.$ Please see the discussion in \cite{ICCpaper, reservoirJSAC}, in particular the comparison between FFT and decimated DFT. Even though decimated DFT improves with increasing decimation up to a threshold $d$, with best performance achieved for $d=8$ for 2 loops, and $d=4$ with 4 loops, FFT matches it with 8 loops and continues to improve up to 10 loops. With fewer loops decimated DFT is better, but we picked FFT as it excels with more loops.% and our loops are inexpensive.

%\begin{figure}[h]
%%\vspace{-1mm}
%\centering
%\hspace{-2mm}\includegraphics[width=0.5\textwidth]{accarch3.png}
%\vspace{-3mm}
%\caption{Accuracy vs. N with decimation of DFT: N= 1200 achieves 97.48\% of accuracy with 2 split loops and $d=8,$ or 4 split loops and $d=4.$
%With further decimation (16-fold) the accuracy falls back to 4-fold decimation (2 loops).
%}\vspace{-5mm}
%\label{fig:accDFT}
%\end{figure}
\section{Discussion of Experimental Results}
\subsection{Basic Performance}
The table of Figures of merit (FOMs) in Fig.~\ref{fig:fom} shows the 20-device SEI performance numbers for the best trade-off we achieved with DLR. These are compared with the equivalents for SoA neural networks trained on a single GPU. The reduction factors {\em (RedF)}, obtained when DLR FOMs are compared to respective values for ResNet or RNN, are: {\em spatial RedF}, the ratio of the number of trainable parameters, is 20, {\em power RF}, the ratio of the complexity of training, is 100, while the latency compares the hours of GPU training versus 5 seconds achieved with our demo platform ($\geq 1200$). For the calculation pattern of FOMs, please see our paper on intermediate results \cite{GomacTech}.The comparison is based on N=600. Note that we could achieve higher accuracy  of above 97\% by using the reservoir with $N=1200,$ which increases the complexity 4$\times$. Fig.~\ref{fig:fomMF} shows the FOMs of MF-DLR for some stationary impairments.
%%% ERASE
%%%%%%%%%%%%%%%%%%%%%%%%%%%%%%%%%%%%
\begin{figure}[h]
\vspace{-2mm}
\centering
%\begin{table}%[]
{\small
\begin{tabular}{l|l|l|l|l}
\hline
 Model  &Accuracy(\%)   &Memory   &Complexity  &Latency  \\
 %& \% & $M$  & $C$ &$\Delta$ \\
 ResNet  & 94.5 & 214K  & $> 5 \times 10^{11}$ &  \\
 LSTM RNN  & 91 & 210K & $ 5 \times 10^{11}$ &  $>1h$\\
 DLR, N=600 & 95  & QN=12K & $5 \times 10^{9}$ & ~5s\\
DLR, N=1K & 96.7  &  20K & $12\times 10^{9}$ & ~20s\\
\hline
\end{tabular}
}
%\end{table}
\vspace{-3mm}
\caption{FOMs compare the best trade-off of {\em SEI accuracy vs H/W reduction} for DLR vs SoA. Complexity is expressed in terms of total MAC operations and memory in terms of the number of parameters to train.
}\vspace{-3mm}
\label{fig:fom}
\end{figure}
\begin{figure}[h]
\vspace{-2mm}
\centering
%\begin{table}%[]
{\small
\begin{tabular}{l|l|l|l}
\hline
 Model  &Accuracy(\%)   &Memory   &Complexity   \\
 %\% & $M$  & $C$ &$\Delta$ \\
 MF-DLR, N=1K & 99.99  & QN=20K & $12 \times 10^{9}$ \\
MF-DLR , N=1K & 99.21  &  20K & $12\times 10^{9}$ \\
AWGN 0dB+jitter& & &\\
MF-DLR , N=1K & 98.61  &  20K & $12\times 10^{9}$ \\
0dB+fading+jitter ($T_4$)& & &\\
\hline
\end{tabular}
}
%\end{table}
%\vspace{-3mm}
\caption{Figures of Merit for MF-DLR on stationary impairments
}\vspace{-4mm}
\label{fig:fomMF}
\end{figure}
\subsection{Robust DLR}
We next address signal corruptions due to receiver imperfections, fading channels, and interference. In order to make SEI robust to these imperfections, we propose a matched filter extension to DLR in its pre-processing stage (MF-DLR).  The term matched filter is somewhat arbitrary as we do not know the waveform shape in the transient turn-on part of the emission. Here is how that {\em matched filer} is implemented: we pick a random datapoint from every device $D$ that we are training on, to be used as the filter for all datapoints of this class, i.e.  applied before we perform the transform (FFT) at the entrance of DLR.  Hence, for 20 devices we have 20 filters. This resulted in a  huge improvement in accuracy for a range of signature corruptions. However, using a random instantiation of signature per device caused randomness in the accuracy across runs. The max variation in accuracy was  ~10\%. We replaced the random instantiation with the average datapoint per device, maintaining robustness with less randomness.

Now, this is somewhat counter-intuitive: to use the right filter we have to know the identity of the device that we are trying to identify. In supervised training we already know the identity, but for inference, we run several filters in parallel (as many as there were devices in the training set).
 If the filter is wrong we will know it by the value of the entropy statistic \cite{reservoirJSAC}, with the false positives of max 5\%  with $T_4$ impairments (see bellow).

For compactness, we emulate the effects of the propagation and the receivers by using 5 distinct compositions of transforms representing fading channels and the complex imperfections in the receiver (RX) front end, including time and frequency jitter, phase noise, I/Q-imbalance, and non-linear power amplification. These transforms are applied to the clear-captured data presented above. We denote classes of these affects as $T_0 \cdots T_4$, where $T_0$ represents clean signals, and classes $T_i$ are characterized by imperfections that increase monotonically with index $i$. $T_i$ parameters dependent on the index define the range of the random effects applied to clean datapoints: $t_i = 25i,$ maximum absolute time error (in samples) in the burst extraction; $p_i=0.25i,$ maximum absolute phase rotation (in radians) of the burst; $f_i = 0.075i$ maximum absolute frequency shift in the burst extraction; $b_i = 0.6-0.1i, $ upper bounds of Rayleigh fading coherence bandwidth, $r_i=1+0.625i$ maximum non-integer resampling rate due to non-linearities in ADC; $s_i=(50-12.5i)dB,$ SNR value. We also model the fading rate and resampling as a linear function of the index. Power-delay profile is fixed to $\set{1,1,1}.$

Finally, with superheterodyne receivers, zero IF receivers have become the main SDR RF front-end. Among the RX effects that may destroy the transmitter RF signature, the IQ imbalance \cite{IQimb}, common to zero IF receivers, is  attributed to the mismatched components of the in-phase (I) and the quadrature (Q) branches, such as an imperfectly balanced local oscillator or baseband low pass filters with mismatched frequency responses. We model the IQ imbalance with $Ia_i=0.75i,$ maximum amplitude IQ imbalance; $Ip_i=0.25i,$ maximum phase IQ imbalance (in degrees); $Id_i=0.025i$ maximum DC bias in  IQ imbalance.%; See figure ()

The use of matched filters in communications is typically limited to cases where the pulse shape is known. The turn-on RF signatures are unknown waveforms. However, when the same receiver is used for the training data collection and the SEI inference, and a clear channel is secured ($T_0$), we observe that the turn-on signature of a device experiences small variations only.  Hence, the average device signature obtained by summation and normalization of all its turn-on bursts seemed like a good model of the constant signature waveform. In addition, statistical models of jitter, channels, and noise are all zero-mean symmetric distributions. This extends to the Rayleigh model, which describes how the complex amplitude fades as the radial component of the sum of two uncorrelated Gaussian random variables. Hence, the matched filter obtained as an average turn-on signature of an augmented $T_i$ dataset is expected to provide a good model of the device under a stationary channel. To see what happens when the channel varies we evaluate when we apply a $T_i$ filter and perform inference on a signal impaired by $T_j, j \neq i.$

\begin{figure}[h]
\vspace{-4mm}
\centering
\hspace{-3mm}\includegraphics[width=0.51\textwidth]{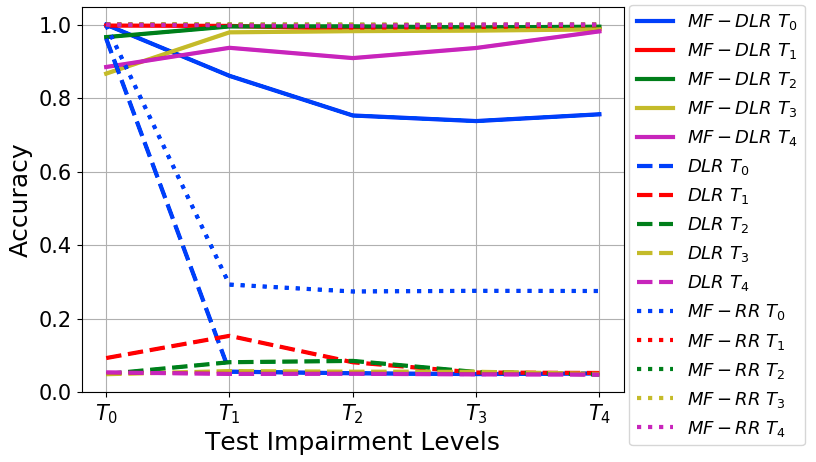}
\vspace{-5mm}
\caption{Accuracy of MF-DLR and MF-RR affected by five classes of imperfections (N=1000, B=600, $\ell = 1024$)
}\vspace{-3mm}
\label{fig:5classes}
\end{figure}

For different choices $T_0 \cdots T_4$ of the training dataset and matched filter (MF), Fig.~\ref{fig:5classes} shows the accuracy as a function of imperfections $T_0 \cdots T_4$ corrupting the evaluation dataset (the test channel). If the matched filter and the training dataset are affected by the same impairments as the evaluation dataset ($T_0 \cdots T_4$), we obtain almost perfect accuracy across channels (with highest 99.99\% with $T_0$ and lowest 99.34 \% for $T_4$). Further we show what happens when the MF and the training set undergo the same specific impairment, while the test-time channel is changing from  $T_0$ to $T_4.$ This is the effect of non-stationarity. Fig.~\ref{fig:5classes} shows how the MF and the delay loop (DLR) affect the robustness of the RR classifier. We see that in the case of the $T_0$ training with MF, delay loop adds additional robustness to changing channels compared to MF-RR (blue curve vs. dotted blue curve). MF-RR appears to be working better for $T_i, i>0$ (by up to 2 \%), which is due to a larger $W$ matrix, and also due to random effects of the random spreading $m(t)$ that was not optimized. Looking at the separation of clean turn-on bursts, we see that they are pretty well separated to start with. Future work will include bursts beyond transients, where we expect increased DL benefits. 

\begin{figure}[h]
\vspace{-1mm}
\centering
\hspace{-6mm}\includegraphics[width=0.35\textwidth]{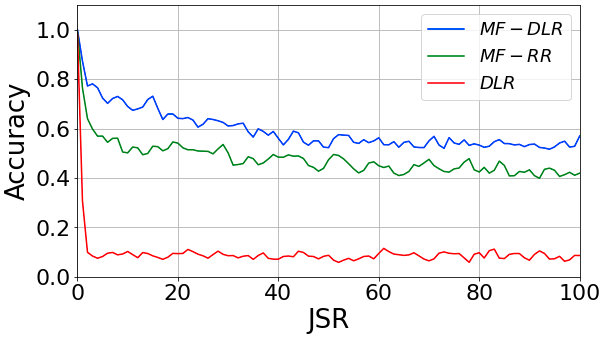}
\vspace{-3mm}
\caption{Accuracy of MF-DLR and MF-RR in the presence of in-band jamming as a function of Jamming to signal ratio (JSR).
}\vspace{-4mm}
\label{fig:inbj}
\end{figure}

%%Jamming
To test the effect of matched filter in the presence of jamming  (in-band interference), we create two subsets from the original WiFi collection. We use the first subset of 12 devices to create the  clean training dataset $\curlb{X_{tr}},$ as well a disjoint dataset of test bursts $\curlb{X_{t}},$ while the second subset of 8 devices $\curlb{X_j}$  was used to create an evaluation set in the following way: we take the burst from the the clean test set $\curlb{X_{t}}$ (at the start of emission), and superpose one or more bursts from $\curlb{X_j}$ (depending on the JSR). Such a 'jammed' datapoint is pushed into DLR, where it is first matched-filtered and then processed as described above. During the training $\curlb{X_{tr}}$ contained 600 bursts for each of 12 'unjammed' devices, and each burst would be match-filtered before entering the reservoir. For  the evaluation, each of 12 jammed devices also had 600 bursts, hence the accuracy results in Fig.~\ref{fig:inbj} are based on the total of 7200 bursts ($\sim 10\mu$s each).
Note that the accuracy of 99.986  (without jamming) means that only 1 of 7200  was misclassified.

Each of the 7200 bursts is jammed in a systematic way so that jamming-to-signal ratio (JSR) has a desired value $x$, and the SEI accuracy in the presence of such jamming is recorded. Specifically, a random combination of jammers (out of the 8) transmits during the transmission of the legitimate burst of 1024 samples, but each with a random delay 0-512.
We then calculate the total jamming energy across all jammers as compared to the original signal's energy and use this as the JSR value in Fig.~\ref{fig:inbj}.
%Besides emulating a communication jammer, this experiment also models the case of a repeater jammer, common in the attacks on radar.
With the matched filter, the accuracy of MF-DLR on normal  datapoints (without jamming)  went to 99.99\% , up 2\% from a simple DLR on 12 devices.
Note that the MF also benefits MF-RR, but DL adds another $\sim 15\%.$ We expect an optimized MF-DLR to perform even better.  %Note that matched filter equally protects the signature regardless of the 'power' of the jamming signals.(TODO)
%%%%%%%%%%%%%%%%%%%%%%%%%%%%%%%%%%%%
%%%%%%%%%%%%%%%%%%%%%%%%%%%%%%%%%%%%
%%%%%%%%%%%%%%%%%%%%%%%%%%%%%%%%%%%%%%%%%%%%%%%%%%%%%%%%%%%%%%%
\vspace{-1mm}\section{Conclusion}\vspace{-1mm}
We showed that Specific Emitter Identification (SEI) for secure IoT authentication can be trained orders of magnitude faster than SoA on simple platforms,  achieving  remarkable accuracy while maintaining this robustness in the presence of fading and interference.  The emphasis of this work is on using 'matched filters' in the pre-processing stage of two classifiers to eliminate the corruption of the RF signature due to the channel. The matched filter is constructed based on the average signature of a given device. As we do not know the device ahead of time, we apply multiple matched filters in parallel and select the correct one using hypothesis testing. If an incorrect filter was used, a statistical test based on the output of the classifier would indicate it with high confidence. The correctly applied filter not only separate the signal from interference but also makes the SEI robust to fading and other imperfections at the receiver. Interestingly, it increases the accuracy on the clean signals too.
%%%%%%%%%%%%%%%%%%%%%%%%%%%%%%%%%
%\vspace{-2mm}\section*{ACKNOWLEDGMENT}%\vspace{-1mm
\vspace{-4mm}
\bibliographystyle{IEEEtran}%
\bibliography{DLRbasicS}
\end{document}